\documentclass[preprint2,times]{aastex62}
\usepackage{amsmath}

\shorttitle{Polytropic index of sunspot loops}
\shortauthors{Krishna Prasad et al.}

\begin{document}
\title{The polytropic index of solar coronal plasma in sunspot fan loops and its temperature dependence}

\correspondingauthor{S. Krishna Prasad}
\email{krishna.prasad@qub.ac.uk}

\author[0000-0002-0735-4501]{S. Krishna Prasad} 
\affiliation{Astrophysics Research Centre, School of Mathematics and Physics,
Queen's University Belfast, Belfast, BT7 1NN, UK}                              

\author{J. O. Raes}
\affiliation{Centre for mathematical Plasma Astrophysics, KU Leuven,
Celestijnenlaan 200B, 3001 Leuven, Belgium}                                  

\author{T. Van Doorsselaere}
\affiliation{Centre for mathematical Plasma Astrophysics, KU Leuven,
Celestijnenlaan 200B, 3001 Leuven, Belgium}

\author{N. Magyar}
\affiliation{Centre for mathematical Plasma Astrophysics, KU Leuven,
Celestijnenlaan 200B, 3001 Leuven, Belgium}                                  

\author{D. B. Jess} 
\affiliation{Astrophysics Research Centre, School of Mathematics and Physics,
Queen's University Belfast, Belfast, BT7 1NN, UK}                              

\affiliation{Department of Physics and Astronomy, California State University
Northridge, Northridge, CA 91330, U.S.A.}                                  

\begin{abstract}
Observations of slow magneto-acoustic waves have been demonstrated to possess a number of applications in coronal seismology. Determination of the polytropic index ($\gamma$) is one such important application. Analysing the amplitudes of oscillations in temperature and density corresponding to a slow magneto-acoustic wave, the polytropic index in the solar corona has been calculated and based on the obtained value it has been inferred that thermal conduction is highly suppressed in a very hot loop in contrast to an earlier report of high thermal conduction in a relatively colder loop. In this study, using SDO/AIA data, we analysed slow magneto-acoustic waves propagating along sunspot fan loops from 30 different active regions and computed polytropic indices for several loops at multiple spatial positions. The obtained $\gamma$ values vary from 1.04$\pm$0.01 to 1.58$\pm$0.12 and most importantly display a temperature dependence indicating higher $\gamma$ at hotter temperatures. This behaviour brings both the previous studies to agreement and perhaps implies a gradual suppression of thermal conduction with increase in temperature of the loop. The observed phase shifts between temperature and density oscillations, however, are substantially larger than that expected from a classical thermal conduction and appear to be influenced by a line-of-sight integration effect on the emission measure. 
\end{abstract}

\keywords{magnetohydrodynamics (MHD) --- methods: observational --- Sun:corona --- Sun: fundamental parameters}

\section{Introduction}
Slow magneto-acoustic waves have been regularly observed in the solar corona since their initial discovery in polar plumes \citep{1997ApJ...491L.111O,1998ApJ...501L.217D,1999ApJ...514..441O} and coronal loops \citep{bc1999SoPh186,2000A&A...355L..23D}. Both propagating and standing versions of these waves have been found and their properties have been extensively studied using observations and theoretical modelling (see review articles by \citealt{2009SSRv..149...65D} and \citealt{2011SSRv..158..397W}). Standing slow waves are mainly observed in flare-related hot loop structures (\citealt{2002ApJ...574L.101W,2005A&A...435..753W,2015ApJ...804....4K,2016ApJ...828...72M,2017A&A...600A..37N}, see \citealt{2017ApJ...847L...5P} for an exception) and are relatively rare whereas propagating slow waves have a photospheric source \citep{2006ApJ...643..540M,2012ApJ...757..160J,2015ApJ...812L..15K} and are more common and ubiquitous in warm loops \citep{2002SoPh..209...61D,2006A&A...448..763M,2012SoPh..279..427K,2012A&A...546A..50K,2014ApJ...789..118K}. Modern high-resolution observations have significantly improved our knowledge on slow waves and additionally revealed a wealth of seismological applications.

Using stereoscopic observations of slow waves from STEREO/EUVI, \citet{2009ApJ...697.1674M} estimated their true propagation speed and thereby deduced the temperature of the associated loop. \citet{2009A&A...503L..25W}, employing spectroscopic observations of slow waves from EUV Imaging Spectrometer (EIS) on board {\it Hinode} \citep{2007SoPh..243...19C}, obtained the inclination angle of a loop in addition to the corresponding plasma temperature. \citet{2011ApJ...727L..32V} reported the measurement of polytropic index in the solar corona for the first time from the observations of slow waves using spectroscopic data from EIS. \citet{2015ApJ...811L..13W} made a similar measurement for the plasma in a hot flare coronal loop utilising the observations of standing slow waves. Applying the dependence of the slow wave propagation speed on the magnetic field for high plasma-$\beta$ loops, the coronal magnetic field has been estimated from both standing \citep{2007ApJ...656..598W} and propagating waves \citep{2016NatPh..12..179J}. Utilising the observations of accelerating slow magneto-acoustic waves in multiple channels, \citet{2017ApJ...834..103K} obtained the spatial variation of temperature along a coronal loop in addition to revealing its underlying multi-thermal structure \citep{2003A&A...404L...1K}.

Our main focus in this study, is however, on the determination of polytropic index. There have been several studies in the past on the estimation of polytropic index using solar wind properties \citep[\textit{e.g.}][]{1963idp..book.....P, 1969SoPh....7..448R, 2006JGRA..11110107K}, but our emphasis here is on the particular application of the observations of slow magneto-acoustic waves. It has been shown that thermal conduction introduces a phase lag between temperature and density perturbations of a slow magneto-acoustic wave \citep{2009A&A...494..339O}. Also, from simple linearised MHD theory for slow waves, one can show that the relative amplitudes of perturbations in temperature and density are directly related through the polytropic index \citep{2003ASSL..294.....G}. Applying these, \citet{2011ApJ...727L..32V} derived a polytropic index, $\gamma$=1.10$\pm$0.02 and inferred that thermal conduction is very efficient in the solar corona. The authors obtained the required temperature and density information from spectroscopic line ratios. Furthermore, through a comparison between temperature and magnetic field fluctuations, found from spectropolarimetric inversions of upper-chromospheric sunspot observations, \citet{2018ApJ...860...28H} also uncovered a similar polytropic index, $\gamma$=1.12$\pm$0.01. \citet{2015ApJ...811L..13W}, on the other hand, employed a differential emission measure analysis on broadband imaging observations of a hot flare loop exhibiting standing slow magneto-acoustic oscillations to obtain a $\gamma$ value of 1.64$\pm$0.08, close to the adiabatic index (5/3). This implies that thermal conduction is highly suppressed in this loop in contrast to the results of \citet{2011ApJ...727L..32V} and \citet{2018ApJ...860...28H}. Extending this work, \citet{2018ApJ...860..107W} performed 1D MHD simulations to compare with the observations and extract further information on excitation and damping mechanisms of slow waves. In this study, we follow the approach of \citet{2015ApJ...811L..13W} and analyse propagating slow magneto-acoustic waves in different active region fan loops using multi-band imaging observations. The details of observations, obtained results, and conclusions are described in the following sections.

\section{Observations}
Fan-like coronal loops rooted in sunspots are selected from 30 different active regions observed between 2011 and 2016, for the present study. Imaging sequences of one-hour long durations taken in 6 coronal channels, namely the 94{\,}\AA, 131{\,}\AA, 171{\,}\AA, 193{\,}\AA, 211{\,}\AA, and 335{\,}{\AA} channels of the Atmospheric Imaging Assembly \citep[AIA;][]{2012SoPh..275...17L} on-board the Solar Dynamics Observatory \citep[SDO;][]{2012SoPh..275....3P} are particularly utilised. AIA cutout data with subfields of about 180$\arcsec\times$180$\arcsec$ encompassing the individual fan-loop structures were obtained and processed for all the 6 channels using a robust pipeline developed by Rob Rutten in IDL\footnote{http://www.staff.science.uu.nl/~rutte101/rridl/sdolib/}. Besides applying the necessary roll angle and plate scale corrections to the downloaded level 1.0 data using \texttt{aia\_prep.pro} (bringing them to science-grade level 1.5), this pipeline aligns images from multiple channels and corrects for any time-dependent shifts using a large subfield disk centre data obtained at a lower cadence. Sub-pixel alignment accuracies of about 0$\farcs$1 are typically achieved even for the target subfields away from disk centre. The spatial and temporal resolutions of the final data are about 0$\farcs$6 and 12{\,}s, respectively. 
\begin{figure*}
\centering
\includegraphics[width=0.95\textwidth, clip=true]{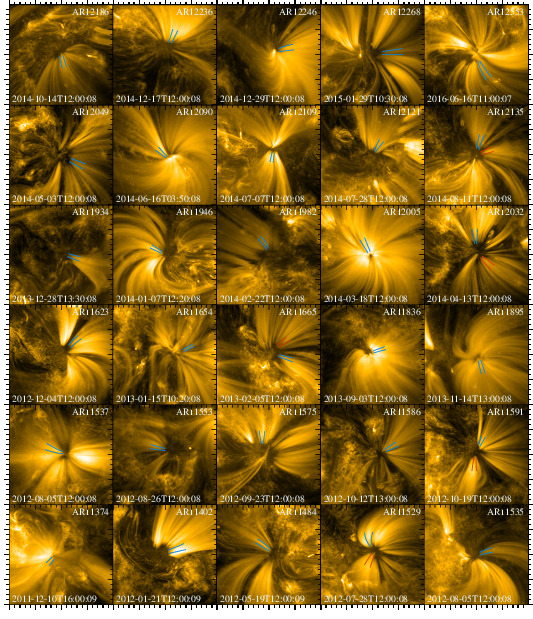} 
\caption{AIA 171{\,}{\AA} images displaying fan-like loop structures from 30 different active regions observed between 2011 and 2016. The respective start times and NOAA numbers are printed on the figure. The solid lines in blue and red represent selected loop segments from individual regions. Blue and red segments correspond to `loop1', and `loop2', respectively, as listed in Table{\,}\ref{tab1}.}
\label{fig1}
\end{figure*}
Fig.{\,}\ref{fig1} displays the vicinity of the selected fan-loop structures from all the 30 active regions using snapshots from the AIA 171{\,}{\AA} channel. The start times of the individual datasets along with the corresponding NOAA numbers are listed in the figure. The central co-ordinates, the oscillation period, and other important parameters obtained in this study are listed in Table{\,}\ref{tab1}.

\section{Analysis and Results}
Fan-like loop structures from each of the selected active regions were manually inspected for propagating oscillations and a loop segment has been chosen where the oscillations show large amplitudes. These loop segments are shown as solid blue lines in Fig.{\,}\ref{fig1}. In about five cases, we found an additional loop segment displaying oscillations with reasonably good amplitudes. These structures are marked with red solid lines in Fig.{\,}\ref{fig1}. We constructed time-distance maps \citep[\textit{e.g.}][]{bc1999SoPh186, 2000A&A...355L..23D} for all the chosen loop segments following a method similar to that described in \citet{2012SoPh..281...67K}. Although the selection of loop segments was mainly based on data from the AIA 171{\,}{\AA} channel, similar time-distance maps were created for all 6 AIA coronal channels using co-spatial segments. The respective intensities from all the 6 channels were then subjected to the regularised inversion code developed by \citet{2012A&A...539A.146H} to obtain the Differential Emission Measure (DEM) at each spatial and temporal position along the loop structure. 

\begin{figure*}
\centering
\includegraphics[width=0.95\textwidth, clip=true]{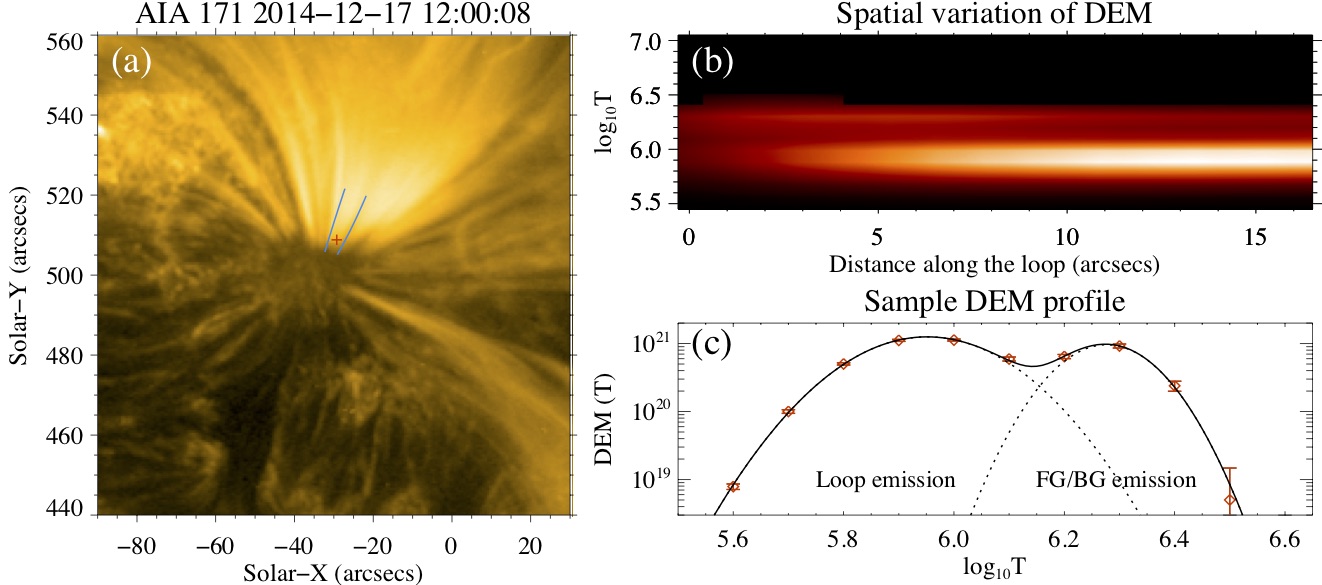} 
\caption{Typical characteristics of obtained DEMs. (a) AIA 171{\,}{\AA} image showing the vicinity of a sunspot fan-loop structure from AR12236 as observed on 2014-12-17, 12:00 UT. The solid blue lines represent the selected loop segment. (b) Time-averaged DEM plot for the loop segment shown in (a) displaying the spatial variation. The double-peaked nature of the DEM is visible throughout the length of the loop with dominant emission from the colder peak. (c) Sample DEM profile from the pixel location marked by a red plus symbol in (a). The solid line represents a double-Gaussian fit to the data in red diamonds. The dotted lines mark the two individual Gaussians corresponding to the emission from the loop and the foreground/background emission.}
\label{fig2}
\end{figure*}
A sample DEM profile is shown in Fig.{\,}\ref{fig2}c. The plus symbol in red along the loop segment shown in Fig.{\,}\ref{fig2}a marks the location from where the sample profile has been extracted. As can be seen, the DEM profile is double peaked with its first peak just under 1 MK (log$_{10}T$=6.0) and second peak near 2 MK (log$_{10}T$=6.3) temperatures. The first peak is relatively stronger and broader which represents the dominant emission coming from the loop whereas the second peak appears to be likely due to the foreground/background emission. Nearly all the loop structures analysed in this study exhibit a similar behaviour. Fig.{\,}\ref{fig2}b shows the temporally averaged DEM depicting its spatial variation along the loop structure. Note that the horizontal axis in this figure shows the distance along the loop while the vertical axis displays the temperature in logarithmic scale. Apparently, the double-peaked behaviour is visible all along the loop and the dominant emission is coming from the low-temperature peak throughout the length except for the bottom few arcseconds where the foreground/background emission dominates. In order to properly isolate the loop emission, we employed a best-fit double-Gaussian model for the DEM profiles. The solid line in Fig.{\,}\ref{fig2}c shows the obtained fit to the data while the dotted lines show individual Gaussians. The temperature at which the first Gaussian peaks, is then considered as a measure of the loop temperature while the area under this curve provides the total emission measure. Following \citet{2013ApJ...778..139S} and \citet{2015ApJ...811L..13W}, we restrict the area measurement to $\pm$2$\sigma$ where $\sigma$ is the width of the Gaussian curve. The emission measure ($EM$) and electron number density ($n$) of the loop are related as $n=\sqrt{\frac{EM}{d}}${\,}, where $d$ is the depth of the loop along the line of sight. Considering a symmetric cross-section for the loop, the depth is then estimated from the width of a Gaussian fitted to the cross-sectional intensity profile of the loop. A suitable location is manually selected along each loop segment where the cross-sectional profile could be better fitted with a Gaussian and the width estimated from that location is considered as the depth of the loop segment throughout its length. As it follows, our main analysis is restricted to a few arcseconds length along each loop structure which makes this approximation reasonable. The density and temperature values thus obtained are used to build time-distance maps in these quantities for each of the selected loop segments. 

\begin{figure*}
\centering
\includegraphics[width=0.95\textwidth, clip=true]{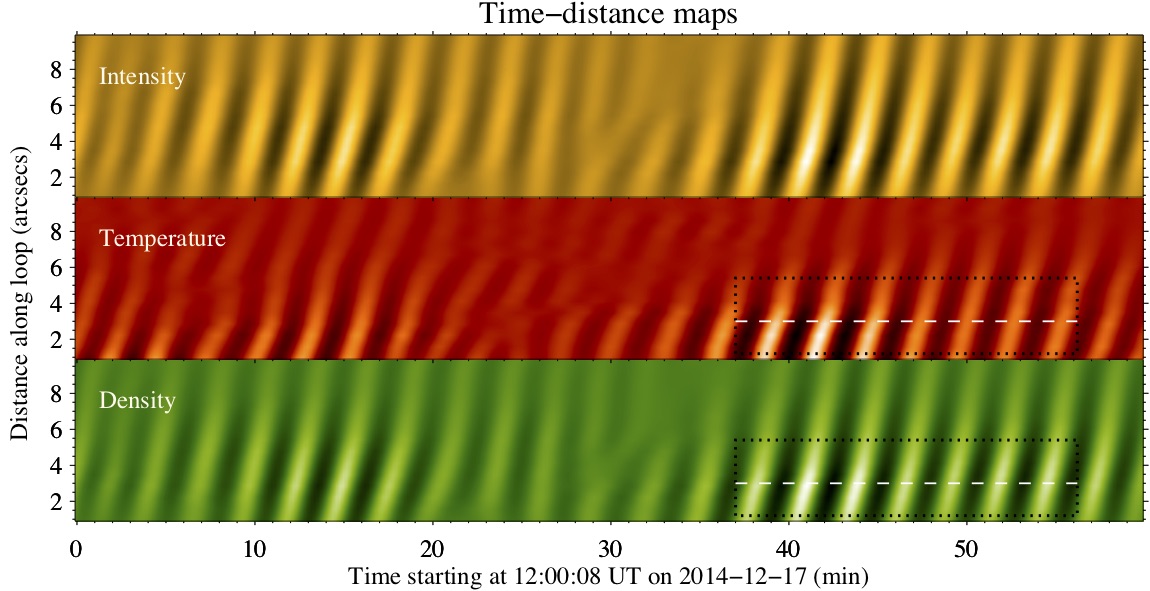} 
\caption{Time-distance maps in AIA 171{\,}{\AA} intensity, temperature, and density, corresponding to the loop segment shown in Fig.{\,}\ref{fig2}a. Each of the maps were Fourier filtered allowing power within a narrow band around the oscillation period to enhance the visibility of the ridges. The boxes in black dotted lines marked on temperature and density maps bound the selected spatial and temporal ranges for phase lag analysis. The white-dashed line in these maps shows the spatial location from which the temperature and density lightcurves plotted in Fig.{\,}\ref{fig4}a are extracted.}
\label{fig3}
\end{figure*}
Sample time-distance maps in intensity, temperature, and density obtained from the loop segment marked in Fig.{\,}\ref{fig2}a are shown in Fig.{\,}\ref{fig3}. The time-distance map in intensity shown here is for the data from the AIA 171{\,}{\AA} channel. The slanted bright/dark ridges in each of these parameters highlight the compressive oscillations propagating along the loop. In order to enhance the visibility of these ridges, the time series at each spatial position has been filtered in the Fourier domain suppressing oscillation power outside a narrow band around the strongest oscillation period. To achieve this, the Fourier power of the respective time series were multiplied by a normalised Gaussian centred at the oscillation period with a width of 1~minute, before applying the inverse Fourier transform that provides the filtered time series \citep[see, \textit{e.g.,}][]{2017ApJ...842...59J}. The oscillation period is pre-determined through a simple Fourier analysis of the AIA 171{\,}{\AA} time series extracted from the average intensities over three adjacent pixel positions close to the bottom of the loop foot point. The average time series has been detrended to remove fluctuations of 6 minutes or longer before establishing the exact oscillation period. 

It may be noted that the compressive oscillations decay rapidly as they propagate along the loop. Therefore, the time-distance maps presented in Fig.{\,}\ref{fig3} are shown only for a small section near the bottom of the loop where the amplitudes are significant. Nevertheless, as can be seen, the temperature perturbations appear to decay faster than those of the density/intensity, a possible consequence of thermal conduction. One may also note that the amplitude of oscillations is not uniform throughout the duration. At certain times (\textit{e.g.,} between 20 to 35 min from the beginning of the time series in Fig.{\,}\ref{fig3}) the oscillation amplitude is very much reduced in all three parameters. Therefore, using the entire time series to determine phase shifts between temperature and density perturbations will produce inaccurate results. Consequently, we restrict the phase difference calculation to a particular range in time and space where the amplitudes are large. This region is shown by the boxes in black dotted lines in Fig.{\,}\ref{fig3}. A similar region has been manually selected for all the loop structures by visually inspecting individual time-distance maps, particularly those of temperature. The key restriction that we imposed while doing this selection is that the region should contain at least three cycles of oscillations in both temperature and density. It may be noted that based on this criterion, we could not find temperature perturbations with sufficient signal in a couple of cases, which, therefore, could not be utilised in further calculations (see Table{\,}\ref{tab1}).

\begin{figure*}
\centering
\includegraphics[width=0.95\textwidth, clip=true]{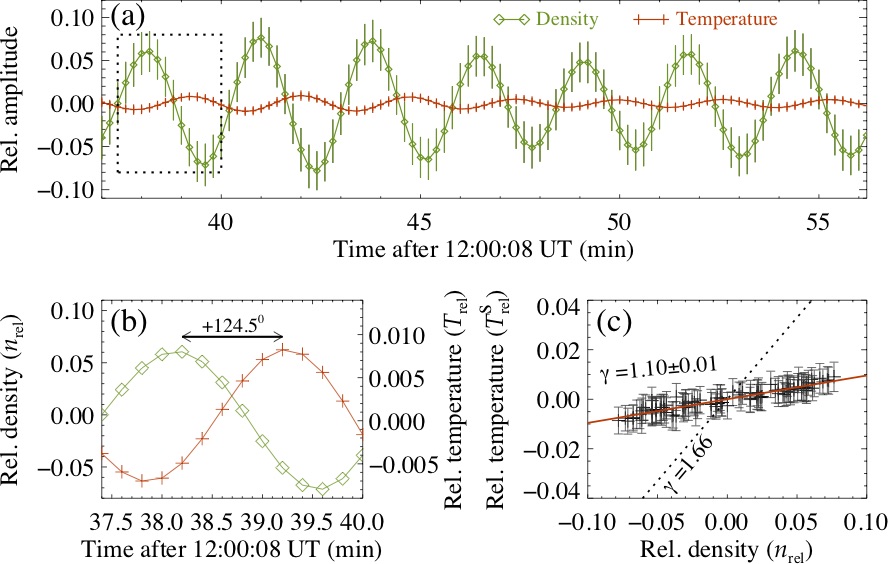} 
\caption{Determination of phase shift and polytropic index. (a) Perturbations in temperature and density from the spatial location marked by the white-dashed lines in Fig.{\,}\ref{fig3}. The vertical bars show the respective uncertainties propagated from the errors on Gaussian fit to DEM curves. (b) A zoomed-in view of the oscillations within the black dotted box in (a) with an independent scale for density and temperature, mainly to highlight the observed phase shift between the quantities. The corresponding uncertainties are not shown here for clarity. (c) A scatter plot of relative temperature (with respective uncertainties as errorbars) plotted against the relative density from (a) after shifting the temperature values to remove the existing phase shift. The red solid line represents the best linear fit to the data. The polytropic index obtained from the slope of the line is listed in the plot. The dotted line shows the expected dependence for adiabatic conditions ($\gamma$=5/3).}
\label{fig4}
\end{figure*}
Fig.{\,}\ref{fig4}a displays relative oscillations in temperature and density within the chosen time range corresponding to the pixel position marked by a white-dashed line in Fig.{\,}\ref{fig3}. Vertical bars, here, denote the uncertainties in respective parameters propagated from the errors on Gaussian fit to DEM curves. It may be noted that the temperature perturbations are considerably smaller than the corresponding density perturbations as one would expect from a linearised MHD theory for slow waves. Using a cross-correlation technique, we measure the time lag ($\Delta$$t$) between the two parameters and then compute the corresponding phase lag ($\Delta\phi$) from it following the relation $\Delta\phi$=($\Delta$$t$/$P$)$\times$360, where $P$ is the oscillation period. The obtained phase difference in this case is about +124$\fdg$5. Fig.{\,}\ref{fig4}b shows a zoomed-in view of the oscillations (within the black dotted box in Fig.{\,}\ref{fig4}a) with a separate scale for temperature and density to clearly highlight the observed phase difference between the quantities. The corresponding uncertainties are not shown here for clarity. 

Following the linearised MHD theory for slow magneto-acoustic waves, including thermal conduction as a damping mechanism \citep[\textit{e.g.,}][]{2003A&A...408..755D, 2009A&A...494..339O, 2014ApJ...789..118K}, it can be shown that 
\begin{eqnarray}
\label{eq1}
T_{rel}\Big(\cos{\,}\Delta\phi &+&\tfrac{\gamma dc_s^2k^2}{\omega}\sin{\,}\Delta\phi-\imath\tfrac{\gamma dc_s^2k^2}{\omega}\cos{\,}\Delta\phi \nonumber \\
&+&\imath\sin{\,}\Delta\phi\Big)=n_{rel}(\gamma-1),
\end{eqnarray}
where $\gamma$ is the polytropic index, $\Delta\phi$ is the phase shift between density and temperature (introduced by thermal conduction), $c_s$=$\sqrt{\gamma p_0/\rho_0}$ is the sound speed, $d$=$\frac{(\gamma-1)k_{\|}T_0}{\gamma c_s^2p_0}$ is the thermal conduction parameter, $\omega$ is the angular frequency, $k$ is the wavenumber, $T_{rel}$=$T^\prime/T_0$, is the relative amplitude of temperature, and $n_{rel}$=$n^\prime/n_0$ is the relative amplitude of density. $n_0$, $T_0$, $\rho_0$, and $p_0$ represent the equilibrium values of electron number density, temperature, mass density, and pressure, respectively. $n^\prime$ and $T^\prime$ denote the corresponding amplitudes of the perturbed values in electron number density and temperature. $k_{\|}$=$k_0T_0^{5/2}$ gives the parallel thermal conduction, where $k_0$ is the thermal conduction coefficient. Equating the imaginary and real parts on both sides of Eq.{\,}\ref{eq1} gives
\begin{eqnarray}
\label{eq2}
\sin{\,}\Delta\phi-\tfrac{\gamma dc_s^2k^2}{\omega}\cos{\,}\Delta\phi = 0, \\
\label{eq3}
T_{rel}\big(\cos{\,}\Delta\phi+\tfrac{\gamma dc_s^2k^2}{\omega}\sin{\,}\Delta\phi\big)=n_{rel}(\gamma-1)
\end{eqnarray}
Using the definition of $d$, and $p_0$=2$n_0k_BT_0$, where $k_B$ is the Boltzmann's constant, Eq.{\,}\ref{eq2} can be rewritten as 
\begin{equation}
\label{eq4}
\tan\Delta\phi = \frac{\pi(\gamma-1)k_{\|}}{k_Bc_s^2Pn_0}
\end{equation}
\citep[\textit{e.g.,}][]{2011ApJ...727L..32V, 2015ApJ...811L..13W}. Here, $P$ is the time period of the oscillation. Additionally, using Eqs.{\,}\ref{eq2} \& \ref{eq3}, one may deduce
\begin{equation}
\label{eq5}
T_{rel}=n_{rel}(\gamma-1)\cos{\,}\Delta\phi.	
\end{equation}
Eqs.{\,}\ref{eq4} \& \ref{eq5} can be used to understand the observed amplitudes of temperature and density, and the phase shifts between them. In the case of fully isothermal plasma, the polytropic index, $\gamma$, is equal to 1 and hence, as may be inferred from Eq.{\,}\ref{eq5}, there wouldn't be any perturbations in temperature. On the other hand, if the conditions are perfectly adiabatic ({\it i.e.,} $\gamma$=5/3) with negligible thermal conduction, Eq.{\,}\ref{eq4} gives $\Delta\phi\approx$0, which implies from Eq.{\,}\ref{eq5} that the relative amplitude of temperature perturbations is about 66\% (2/3) of that of density perturbations with no phase shift between the quantities. For intermediate cases, the knowledge of the polytropic index and thermal conduction coefficient is necessary to accurately determine the values. Alternatively, one can use the observed phase shifts and amplitude values to obtain the polytropic index and the thermal conduction coefficient.

As one might note from Eqs.{\,}\ref{eq4} \& \ref{eq5}, a valid common solution only exists if the phase shift between temperature and density is between 0$\degr$ and 90$\degr$. However, as can be seen from Fig.{\,}\ref{fig4}b the observed phase shift is about +124$\fdg$5, albeit for a single spatial position along the loop structure shown in Fig.{\,}\ref{fig2}a. In fact, even other spatial positions and nearly all the loop structures investigated in this study show similar values well outside the expected range. This implies there is an additional source of phase shift in observations and consequently, the observed values cannot be directly used to obtain the thermodynamic parameters. But, considering the phase shift to be constant over the duration of the time series\footnote{This is a valid assumption since the physical conditions on which the phase shift is dependent upon (see Eq.{\,}\ref{eq4}), do not change appreciably over the time scales involved.}, one can eliminate the dependence on it from Eq.{\,}\ref{eq5} by merely shifting the temperature (or density) time series to match the phase with that of density (or temperature). This step reduces Eq.{\,}\ref{eq5} to 
\begin{equation}
\label{eq6}
T_{rel}^s=n_{rel}(\gamma-1),
\end{equation}
where $T_{rel}^s$ denotes the relative amplitudes of temperature shifted to be in phase with density. Eq.{\,}\ref{eq6} can then be readily used to obtain the polytropic index. In Fig.{\,}\ref{fig4}c, we plot $T_{rel}^s$ obtained by shifting the temperature perturbations shown in Fig.{\,}\ref{fig4}a by $-$124$\fdg$5 against the corresponding unshifted $n_{rel}$. The error bars here denote the respective uncertainties in temperature. A linear relationship between the parameters is evident from the data. The overplotted red solid line denotes the best linear fit obtained. Applying Eq.{\,}\ref{eq6}, we calculate the polytropic index, $\gamma$=1.10$\pm$0.01 from the slope of the fitted line. The expected dependence for adiabatic conditions ($\gamma$=5/3) is shown by a dotted line in this figure, which is largely deviated from the actual data.

It is important to note here that Eq.{\,}\ref{eq6} is generally applicable for slow magneto-acoustic waves whether or not there is thermal conduction. But, of course, in case of no thermal conduction, the temperature perturbations need not be shifted as they are expected to be already in phase with density. However, it does not mean that by removing the phase shift dependence from Eq.{\,}\ref{eq5}, we have completely eliminated the effects of thermal conduction. It is inherently assumed that the polytropic index, $\gamma$, is modified by the presence of thermal conduction which should be reflected in the temperature and density amplitudes of a slow wave. Although this is true, in the case of coronal plasma, the polytropic index is governed not just by the thermal conduction but also by several other important processes such as heating, radiative losses, turbulence, plasma flows and other non thermal processes. Therefore, a $\gamma$ value lower than 5/3, as deduced from the observed amplitudes, need not necessarily imply enhanced thermal conduction.

\begin{figure*}
\centering
\includegraphics[width=0.8\textwidth, clip=true]{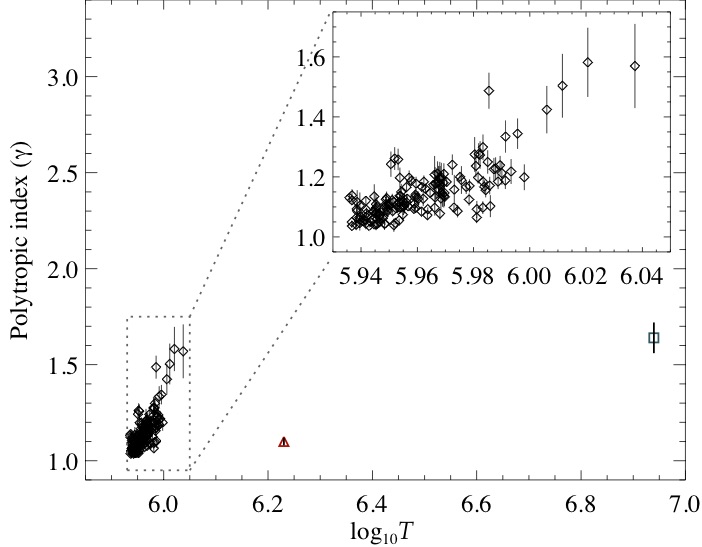} 
\caption{Dependence of polytropic index on mean temperature. Here, $\gamma$ values obtained from individual spatial positions across all the selected loop segments are plotted against the corresponding time-averaged temperature. The diamond symbols in black represent the data from this study whereas the red triangle, and the green square, represent the data from \citet{2011ApJ...727L..32V}, and \citet{2015ApJ...811L..13W}, respectively. The vertical bars show the respective uncertainties on the values. A zoomed-in view of our data is provided in the inset panel to clearly highlight the observed dependence.}
\label{fig5}
\end{figure*}
Keeping the above limitations in mind, we computed the polytropic index at individual spatial positions across all the selected loop structures following the same procedure. The obtained values are plotted against the corresponding time-averaged temperature (logarithmic values) in Fig.{\,}\ref{fig5}, using diamond symbols in black. For comparison, we also show the results from \citet{2011ApJ...727L..32V}, and \citet{2015ApJ...811L..13W}, using a red triangle, and a green square, respectively. The vertical bars in this figure represent the respective uncertainties. All our data, across 164 spatial positions identified from about 33 loop structures, are displayed in this figure. To clearly highlight the observed dependence from our data, a zoomed-in view is presented in the inset panel. Although our temperature range is limited due to our selection of only fan-like warm loop structures, it appears that the polytropic index is increasing with the temperature.

\begin{table*}
\footnotesize
\begin{center}
\caption{Mean values of polytropic index, phase shift, and other important parameters for individual loop structures.}
\label{tab1}
\begin{tabular}{ccccccccccc}
\hline\hline
Start time &NOAA & Location & Loop & $P$ & $A_{I}$ & $T_0$ & $n_0$ & $\gamma$ & $\Delta\phi_{obs}$ & $\Delta\phi_{S}$ \\
(UT) &($\#$) &($\arcsec$) &($\#$) &(s) & &(MK) &($\times$10$^{9}$ cm$^{-3}$) & &($\degr$) &($\degr$) \\
\hline
2011-12-10 16:00&11374&(-663,-281)&loop1&167&0.01&0.90$\pm$0.01&2.9$\pm$0.1&1.22$\pm$0.05& 52.3$\pm$ 9.5&  2.3\\ 
2012-01-21 12:00&11402&(46,549)   &loop1&182&0.05&0.89$\pm$0.01&2.5$\pm$0.2&1.04$\pm$0.01&170.3$\pm$ 6.6&  0.6\\
2012-05-19 12:00&11484&(-136,208) &loop1&198&0.02&0.88$\pm$0.01&2.0$\pm$0.1&1.11$\pm$0.02&204.5$\pm$ 3.9&  1.5\\
2012-07-28 12:00&11529&(-178,-269)&loop1&236&0.02&0.88$\pm$0.01&1.3$\pm$0.1&1.04$\pm$0.01&160.7$\pm$ 7.2&  0.8\\
			    &     &           &loop2&153&0.05&0.90$\pm$0.01&1.3$\pm$0.1&1.12$\pm$0.02&170.6$\pm$ 4.0&  3.5\\
2012-08-05 12:00&11535&(172,221)  &loop1&182&0.03&0.93$\pm$0.01&0.8$\pm$0.1&1.20$\pm$0.01&183.3$\pm$14.8&  7.2\\
2012-08-05 12:00&11537&(-269,115) &loop1&167&0.03&0.87$\pm$0.01&1.2$\pm$0.1&1.08$\pm$0.03&198.1$\pm$19.8&  2.3\\
2012-08-26 12:00&11553&(-428,-433)&loop1&182&0.03&0.97$\pm$0.01&1.6$\pm$0.1&1.37$\pm$0.11&185.7$\pm$16.0&  6.3\\
2012-09-23 12:00&11575&(-215,15)  &loop1&236&0.01&0.91$\pm$0.02&1.5$\pm$0.2&1.12$\pm$0.02&218.9$\pm$14.5&  2.1\\
2012-10-12 13:00&11586&(-83,-303) &loop1&182&0.03&0.90$\pm$0.01&1.4$\pm$0.1&1.10$\pm$0.02&195.6$\pm$19.4&  2.2\\
2012-10-19 12:00&11591&(240,30)   &loop1&167&0.09&0.87$\pm$0.01&1.7$\pm$0.1&1.05$\pm$0.01&201.9$\pm$ 1.4&  1.0\\
			    &     &           &loop2&182&0.04&0.91$\pm$0.01&1.4$\pm$0.2&1.09$\pm$0.01&163.1$\pm$10.7&  2.1\\
2012-12-04 12:00&11623&(230,116)  &loop1&167&0.04&0.91$\pm$0.01&1.5$\pm$0.1&1.11$\pm$0.01&168.4$\pm$ 2.1&  2.6\\
2013-01-15 10:20&11654&(310,176)  &loop1&140&0.04&0.96$\pm$0.01&1.8$\pm$0.1&1.19$\pm$0.03&186.2$\pm$ 6.8&  4.5\\
2013-02-05 12:00&11665&(355,289)  &loop1&182&0.06&0.97$\pm$0.01&1.1$\pm$0.1&1.16$\pm$0.03&192.0$\pm$ 3.5&  5.0\\
				&	  & 	          &loop2&182&0.03&0.97$\pm$0.01&0.9$\pm$0.1&1.25$\pm$0.03&168.8$\pm$ 3.8&  8.3\\
2013-09-03 12:00&11836&(276,70)   &loop1&167&0.03&0.90$\pm$0.01&3.8$\pm$0.1&1.11$\pm$0.01&105.2$\pm$ 5.9&  1.0\\
2013-11-14 13:00&11895&(-278,-316)&loop1&167&0.02&0.89$\pm$0.01&2.3$\pm$0.1&1.12$\pm$0.01&164.1$\pm$ 5.9&  1.7\\
2013-12-28 13:30&11934&(576,-215) &loop1&140&0.03&0.96$\pm$0.01&1.4$\pm$0.1&1.11$\pm$0.05&169.8$\pm$25.9&  3.6\\
2014-01-07 12:20&11946&(-96,220)  &loop1&167&0.02&0.88$\pm$0.01&2.4$\pm$0.3&\nodata&\nodata&\nodata\\
2014-02-22 12:00&11982&(-252,-38) &loop1&182&0.02&0.93$\pm$0.01&2.0$\pm$0.2&1.16$\pm$0.03&179.8$\pm$ 5.1&  2.5\\
2014-03-18 12:00&12005&(13,317)   &loop1&167&0.01&0.91$\pm$0.01&4.1$\pm$0.1&1.15$\pm$0.02&149.6$\pm$ 3.1&  1.2\\
2014-04-13 12:00&12032&(-68,280)  &loop1&257&0.01&0.95$\pm$0.02&1.2$\pm$0.2&1.09$\pm$0.01&197.2$\pm$ 6.7&  1.8\\
				&	  &			  &loop2&182&0.03&0.87$\pm$0.01&1.8$\pm$0.1&1.11$\pm$0.04&117.9$\pm$ 5.7&  1.8\\
2014-05-03 12:00&12049&(48,-80)   &loop1&198&0.02&0.93$\pm$0.01&1.2$\pm$0.1&1.17$\pm$0.02&151.3$\pm$ 4.3&  4.1\\
2014-06-16 03:50&12090&(-196,380) &loop1&182&0.02&0.94$\pm$0.02&2.9$\pm$0.2&\nodata&\nodata&\nodata\\
2014-07-07 12:00&12109&(-209,-193)&loop1&153&0.07&0.89$\pm$0.01&3.2$\pm$0.2&1.06$\pm$0.01&161.8$\pm$15.4&  0.8\\
2014-07-28 12:00&12121&(87,40)    &loop1&167&0.01&0.88$\pm$0.01&2.1$\pm$0.1&1.08$\pm$0.02&135.4$\pm$10.8&  1.3\\
2014-08-11 12:00&12135&(33,132)   &loop1&153&0.01&0.97$\pm$0.03&1.3$\pm$0.1&1.29$\pm$0.08&187.1$\pm$11.6&  7.8\\
				&	  &			  &loop2&167&0.02&1.06$\pm$0.03&1.1$\pm$0.1&1.55$\pm$0.04&189.2$\pm$ 3.8& 14.7\\
2014-10-14 12:00&12186&(166,-436) &loop1&182&0.02&0.88$\pm$0.01&2.3$\pm$0.1&1.07$\pm$0.01&132.1$\pm$16.9&  1.0\\
2014-12-17 12:00&12236&(-35,492)  &loop1&167&0.04&0.90$\pm$0.02&1.6$\pm$0.3&1.09$\pm$0.01&124.3$\pm$ 8.3&  2.1\\
2014-12-29 12:00&12246&(347,330)  &loop1&182&0.04&0.92$\pm$0.01&1.3$\pm$0.1&1.16$\pm$0.02&171.2$\pm$ 8.5&  3.8\\
2015-01-29 10:30&12268&(275,-60)  &loop1&167&0.04&0.87$\pm$0.01&1.4$\pm$0.1&1.06$\pm$0.01&171.4$\pm$ 9.5&  1.4\\
2016-06-16 11:00&12553&(11,-129)  &loop1&182&0.03&0.95$\pm$0.03&2.1$\pm$0.1&1.19$\pm$0.02&180.5$\pm$21.5&  2.9\\ \hline
\end{tabular}
\tablecomments{`loop1' and `loop2' are shown with blue and red solid lines, respectively, in Fig.{\,}\ref{fig1}. $A_{I}$ denotes the relative intensity amplitudes in AIA 171{\,}{\AA} channel, measured at the bottom of the selected loop segments.}
\end{center}
\end{table*}
In Table{\,}\ref{tab1}, we list the mean values of temperature, $T_0$, density, $n_0$, the polytropic index, $\gamma$, and the observed phase shift between the temperature and density perturbations ($\Delta\phi_{obs}$) for all the loop structures studied. The uncertainties listed with these values are obtained from the respective standard deviations across each loop. The small uncertainties suggest that the values themselves do not vary much within a loop structure. The limited spatial ranges considered along each loop could also be partially responsible for this. It may be noted that the temperature and density values obtained here are of the same order of those found from spectroscopic observations of fan loops \citep{2017ApJ...835..244G}. The relative intensity amplitudes and the oscillation periods computed for individual loops, are also listed. The amplitudes are measured from the AIA 171{\,}{\AA} channel, near the bottom of the selected loop structures where they are usually the highest. The values range from 0.01 to 0.09 suggesting the linear nature of the observed waves. A majority of the oscillation periods are near 180{\,}s, which is not surprising considering the fact that the selected loop structures are rooted mainly in sunspots. Additionally, in the last column, we list the expected phase shifts ($\Delta\phi_S$) obtained using Eq.{\,}\ref{eq4}, for a classical Spitzer thermal conductivity \citep[$k_0$=7.8$\times$10$^{-7}$;][]{1962pfig.book.....S}. The polytropic index, temperature, density, and the oscillation period used in this calculation are from the observed values. While the values $\Delta\phi_S$ range between 0$\fdg$5 to 15$\degr$, the observed phase shifts, $\Delta\phi_{obs}$ vary between 52$\degr$ to 219$\degr$. Evidently, the observed phase shifts are significantly larger than the corresponding values for classical thermal conductivity. In a couple of loops, the temperature perturbations do not possess sufficient amplitudes to estimate $\gamma$, $\Delta\phi_{obs}$, and $\Delta\phi_S$, which therefore, are left blank.

\section{Discussion and Conclusions}
It has been demonstrated that the amplitudes of temperature and density perturbations due to a slow magneto-acoustic wave, and the phase shift introduced between them by thermal conduction can be utilised to understand the thermodynamic properties of solar coronal plasma. Using spectroscopic data from Hinode/EIS, \citet{2011ApJ...727L..32V} have obtained the polytropic index, $\gamma$=1.10$\pm$0.02 for a warm coronal loop and inferred that thermal conduction is very efficient in the solar corona. More recently, \citet{2015ApJ...811L..13W} have performed a similar analysis on a hot flare loop and obtained $\gamma$=1.64$\pm$0.08 suggesting, in contrast, a suppression in thermal conduction. Although the relevant spectroscopic data were not available, \citet{2015ApJ...811L..13W} have extracted the required temperature and density information from broadband SDO/AIA images in multiple coronal channels through DEM analysis. Following the latter approach, we studied propagating slow magneto-acoustic waves in about 30 different active regions, particularly those in fan-like loop structures. Employing a regularised inversion method \citep{2012A&A...539A.146H} on the observed intensities in 6 AIA coronal channels, corresponding DEMs have been obtained. The DEM profiles mostly displayed a double-peaked structure with one broader dominant peak around 1MK representing the loop emission and another narrow peak near 2 MK representing the foreground/background emission. We carefully isolated the loop emission using the best-fit double-Gaussian profiles and computed respective electron temperatures and densities (via emission measure) to construct time-distance maps in those parameters along selected loop segments. For each loop structure, a limited range in temporal and spatial domains is identified near the bottom of the loop where the amplitudes of oscillation are large enough to accurately calculate the phase shift between temperature and density. Within the selected range, the phase shift between the parameters is computed using a cross-correlation method. The obtained values are substantially larger than that expected from a classical Spitzer thermal conductivity. While this might mean that the thermal conduction is higher than the classical values, one must note that the observed values are even larger than that could be reconciled with the wave theory, which therefore suggests that the apparent phase shifts are not merely due to thermal conduction.

It is also worth noting that most of the observed phase shift values are clustered around 180$\degr$ (see Table{\,}\ref{tab1}). \citet{2012ApJ...757..160J} studied a sunspot fan-loop structure using an independent DEM method and found that the obtained peak temperature and emission measure are 180$\degr$ out of phase. The authors explained this behaviour as due to the anti-correlated changes in emission volume along the line of sight. Although, from our data, we do not find a definitive dependence of phase shifts on the location of the loop structure (perhaps because of the non-uniform distribution of our data with more samples towards solar disk center), it is possible that a similar effect is responsible for the large phase shifts observed here. For instance, the loop structure that is furthest from the disk centre (\textit{i.e.,} from AR11374) in our sample shows the smallest phase shifts ($\approx$52$\degr$). Besides, the temperature and density values obtained from our DEM analysis are representative of mean values over the cross-section of a loop which implies the multi-thermal nature of the active region loops, as evidenced from the differential propagation of slow waves when observed in multiple temperature channels \citep{2012SoPh..279..427K, 2013ApJ...778...26U, 2017ApJ...834..103K}, is not considered. This approximation may as well influence the observed phase shifts. It may also be worthwhile exploring if the misbalance in the local thermal equilibrium caused by the slow waves \citep{2017ApJ...849...62N} can in turn affect the observed phase shifts. Additional effects, such as nonlinearity \citep{2000A&A...362.1151N, 2002ApJ...580L..85O} and partial wave reflection can impact the phase shifts but do not seem to be applicable to our data. In any case, even in the absence of above effects, the exact phase shift depends on several factors (see Eq.{\,}\ref{eq4}) and hence, should not be assumed to correlate precisely with thermal conduction.

By manually shifting the temperature time series to remove the existing phase shift, we compare the oscillation amplitudes in temperature against that in density and compute the polytropic index. These values occur in the range between 1.04$\pm$0.01 and 1.58$\pm$0.12. It also appears that there is a temperature dependence with hotter loops having higher polytropic index. Qualitatively, this behaviour brings the contrasting findings of \citet{2011ApJ...727L..32V} and \citet{2015ApJ...811L..13W} to a good agreement. One may note, however, that the observed dependence of polytropic index on temperature (see Fig.{\,}\ref{fig5}) is steeper than that would be required to have a better match with the previous results. This warrants the requirement of additional examples distributed across a wider temperature range to find the exact dependence. Moreover, the phase shift between the temperature and density perturbations observed in the earlier studies is relatively small ($<$90$\degr$) unlike that in our case. \citet{2011ApJ...727L..32V} obtained the temperature and density information directly from the spectroscopic line ratios, so their results are less prone to the line-of-sight effect which we believe is the cause for the large phase shifts in our data. But, it is intriguing why \citet{2015ApJ...811L..13W} do not see any such effect. It is possible that the extremely hot plasma in their loop somehow helps in mitigating the line-of-sight changes, but this requires further studies to confirm and improve our understanding of this effect. 

Finally, while it is possible that the increase in polytropic index with temperature might imply a gradual suppression of thermal conduction in agreement with the inferences from previous studies, it is much harder to explain. \citet{2015ApJ...811L..13W} offered some explanations based on nonlocal conduction, plasma waves, and turbulence that are more applicable to hot flare loops but since this behaviour appears to be more prevalent even in the warm loops, one needs to find a general theory. Besides, the influence of other thermodynamic processes such as heating, radiative losses etc., in addition to the observational effects (\textit{e.g.,} multi-thermal structure of loops), on the polytropic index, should be investigated. We believe that future studies, ideally a combination of observations, numerical simulations, and forward modelling, might reveal important information to address this problem. 

\acknowledgements 
The authors thank the anonymous referee for useful comments. SKP is grateful to the UK Science and Technology Facilities Council (STFC) for funding support that allowed this project to be undertaken. TVD was supported by the GOA-2015-014 (KU~Leuven) and the European Research Council (ERC) under the European Union's Horizon 2020 research and innovation programme (grant agreement No 724326). DBJ would like to thank STFC for an Ernest Rutherford Fellowship, in addition to Invest NI and Randox Laboratories Ltd. for the award of a Research \& Development Grant (059RDEN-1). AIA data used here are courtesy of NASA/SDO and the AIA science team. We acknowledge the use of pipeline developed by Rob Rutten to extract, process, and co-align AIA cutout data.


\end{document}